\documentclass{ws-ijmpa}

\begin{document}

\markboth{D.Yu. Bogachev, A.V. Gladyshev, D.I. Kazakov, A.S. Nechaev}
{Light Superpartners at Hadron Colliders}

%
\catchline{}{}{}{}{}
%

\title{LIGHT SUPERPARTNERS AT HADRON COLLIDERS}

\author{D.YU. BOGACHEV$^{2}$, A.V. GLADYSHEV$^{1,2,}$\footnote{e-mail: gladysh@theor.jinr.ru} ,
D.I. KAZAKOV$^{1,2,}$\footnote{e-mail: kazakovd@theor.jinr.ru}
\ and A.S. NECHAEV$^{2}$}

\address{}
\address{$^{1}$Bogoliubov Laboratory of
Theoretical Physics,
Joint Institute for Nuclear Research, \\
141980, 6 Joliot-Curie, Dubna, Moscow Region, Russian
Federation}

\address{$^{2}$Institute for Theoretical and Experimental Physics, \\
117218, 25 B.Cheremushkinskaya, Moscow, Russian Federation}

\maketitle

\begin{history}
\received{Day Month Year}
\revised{Day Month Year}
\end{history}

\begin{abstract}
Uncertainties of the MSSM predictions are due to an unknown
SUSY breaking mechanism. To reduce these uncertainties, one
usually imposes constraints on the MSSM parameter space.
Recently, two new constraints became available, both from
astrophysics: WMAP precise measurement of the amount of the
Dark Matter in the Universe and EGRET data on an excess in
diffuse gamma ray flux. Being interpreted as a manifestation
of supersymmetry these data lead to severe constraints on
parameter space and single out a very restricted area. The key
feature of this area is the splitting of light gauginos from
heavy squarks and sleptons. We study the phenomenological
properties of this scenario, in particular, the cross-sections
of superparticle production, their decay patterns and
signatures for observation at hadron colliders, Tevatron and
LHC. We found that weakly interacting particles in this area
are very light so that the cross-sections may reach fractions of a
pb with jets and/or leptons as final states accompanied by
missing energy taken away by light neutralino with a mass
around 100 GeV.

\end{abstract}

\ccode{PACS numbers: }

\section{Introduction}
\label{intro}

Search for supersymmetry at accelerators of the previous
decade did not result in discovery of any new physics but
rather pushed forward the boundary of unknown territory up to
a few hundred GeV. Still the low energy supersymmetry and
first of all the MSSM~\cite{MSSM1,MSSM2,MSSM3,MSSM4,MSSM5}
proved to be a consistent
model compatible with all experimental data and promissing new
discoveries round the corner. Numerous attempts to fit the
theoretical and experimental requirements with the MSSM have
led to a consistent picture of restricted parameter space
where prediction of the particle spectra and of the
cross-sections of supersymmetry production is
possible~\cite{CMSSM1,CMSSM2,CMSSM3,CMSSM4,CMSSM5,CMSSM6,CMSSM7,CMSSM8,CMSSM9,CMSSM10,CMSSM11,CMSSM12,CMSSM13}.
Though the details depend on the choice
of constraints and the mechanism of supersymmetry breaking,
the allowed region of parameter space still indicates the
presence of light superpartners within the reach of modern
accelerators.

The main hopes of the last decade were connected with the LEP
II $e^+e^-$ collider, where the light charginos, the
superpartners of the weak gauge bosons and charged Higgses,
might be produced together with light sleptons, the
superpartners of the three generations of leptons. As for the
strongly interacting particles, squarks and gluino, they are
typically much heavier, at least within mSUGRA models, and the
corresponding production cross-sections are suppressed.

The situation has changed after LEP shutdown since now we have
the Tevatron and soon coming LHC hadron colliders. There one
expects that first of all the strongly interacting particles
will be produced because the cross-section is enhanced by the
strong coupling. However, the severe background coming from
the SM particles is essential and one has a problem in
extracting the signal from the background.

Recently, a new ingredient to this scheme came from
astroparticle physics where considerable experimental
activities, mostly in space, has led to remarkably precise
data. In particular, the WMAP collaboration measuring the
thermal fluctuations of the Cosmic Microwave Background
determined the matter content of the Universe resulting in $23
\pm 4$\% attributed to the Dark Matter~\cite{WMAP1,WMAP2}. It
perfectly fits supersymmetry, since SUSY provides an excellent
candidate for the Dark Matter particle, namely, the lightest
neutralino, a mixture of the superpartners of the photon,
\mbox{$Z$-boson} and neutral Higgses. Due to a high precision the
WMAP data provide a very restrictive constraint on SUSY
models, and allow one to further restrict the parameter space.

The other contribution comes from the cosmic ray data. It
refers to the measurement of the flux and the spectrum of the
gamma rays and antiparticles coming to the Earth. In
particular, the diffuse gamma rays were measured in a
satellite experiment by the EGRET collaboration and result in
an excess above the background for the energies above 1
GeV~\cite{EGRET1,EGRET2,EGRET3,EGRET4,EGRET5,EGRET6}.
Being interpreted as additional contribution
coming from the SUSY Dark Matter annihilation it leads to
rather restrictive constraint on the value of the neutralino
mass and, respectively, to the restriction on the parameter
space of the MSSM~\cite{new1,new2,new3,new4,new5,new6}.

We show below that taking into account all these constraints
results in a very narrow region of the parameter space.
Choosing parameters in this region one can calculate the
spectrum of superpartners and the cross-section of their
production. These cross-sections happen to be relatively large
to be of interest for future experiments at hadron colliders.
Moreover, as it will be shown below, the cross-section for
the chargino production given by the weak processes is unexpectedly high
and approaches that for the strong processes of
squark and gluino production. This might serve as a signature
for supersymmetry in the future LHC experiments.

\section{Features of the EGRET preferred region}
\label{sec:egret}

The framework of our analysis is the Minimal Supersymmetric
Standard Model with supergravity inspired supersymmetry
breaking terms. The parameters of the model are those of the
Standard Model (three gauge couplings $\alpha_i$, and three $3
\times 3$ matrices of the Yukawa couplings $y^i_{ab}$, where
$i = L, U, D$.), Higgs mixing parameter  $\mu$, and a set of
SUSY breaking parameters (mass terms for squarks and sleptons,
mass terms for gauginos and bilinear and trilinear terms).

In the general case, the MSSM contains more than a hundred
unknown parameters. Most of them come from the SUSY breaking
sector and are the main source of uncertainties. To reduce the
number of unknown parameters, one usually imposes some
constraints, both simple and obvious, and model-dependent
ones.

One of the strictest constraints is the gauge couplings
unification, it fixes the threshold of supersymmetry breaking
$M_{SUSY} \sim 1$~TeV, thus fixing the scale of superparticle
masses~\cite{Unif1,Unif2,Unif3}. The second very hard constraint follows
from the requirement of radiative electroweak symmetry
breaking. While running from the GUT scale towards the lower
energies, one (or both) of the Higgs mass-squared parameters
$m_{H_i}^2$ becomes negative; thus the condition for
spontaneous breaking of electroweak symmetry is fulfilled.
Requiring the breaking to take place at the EW scale $M_{EW}
\sim 100$~GeV restricts the initial conditions for the
corresponding RGE at the GUT scale for the Higgs masses to be
equal to $\sqrt{m_0^2+\mu^2}$, thus expressing the value of
the $\mu$ parameter in terms of $m_0$ and
$m_{1/2}$. The sign of $\mu$ remains
undetermined. One can fix it from other constraints, for
example, from the anomalous magnetic moment of the muon which
has a small deviation from the Standard Model predictions of
the order of 2$\sigma$. This deficiency may be easily filled
with the SUSY contribution, which is proportional to $\mu$.
This requires a positive sign of $\mu$ that kills a half of
the parameter space of the
MSSM~\cite{Anom1,Anom2,Anom3,Anom4,Anom5,Anom6,Anom7,Anom8}.

Further constraints are due to flavour changing processes like
$b \to s \gamma$ responsible for the rare $B$-meson decays
which can occur at the one-loop level due to a virtual $W$-top
pair and are strongly suppressed in the Standard Model. In the
supersymmetric model contribution to the branching ratio $BR(b
\to s \gamma)$ due to superpartners exchange may be rather
big, exceeding the experimental value by a few standard
deviations. However, the next-to-leading order corrections are
essential and improve the situation. This requirement imposes
severe restrictions on the parameter space, especially in the
case of high
$\tan\beta$~\cite{bsg1,bsg2,bsg3,bsg4,bsg5,bsg6,bsg7,bsg8,bsg9,bsg10,bsg11,bsg12,bsg13,bsg14}.

Experimental lower limits on the Higgs boson mass~\cite{Higgs}
impose further constraints when taking into account two-loop
radiative corrections. This limit ($m_h~\le~113.4$~GeV)
forbids left lower corner of the $m_0 - m_{1/2}$ plane,
together with the $b\to s \gamma$
constraint~\cite{Anom7,higgscon1,higgscon2}.
Conservation of $R$-parity results in the existence of the
lightest supersymmetric particle (LSP) which is usually the
lightest neutralino $\chi^0_1$. The neutralino is a perfect
candidate for the non-baryonic cold Dark Matter particle. The
requirement that LSP is the lightest neutralino excludes the
whole area in the parameter space, where LSP is the charged
stau~\cite{LSP1,LSP2}.

Recent very precise data from WMAP collaboration, which
measured thermal fluctuations of Cosmic Microwave Background
radiation, restricted the amount of the Dark matter in the
Universe up to $23 \pm 4\%$.  This imposes further
restrictions on the model. In the early Universe all particles
were produced abundantly and were in thermal equilibrium
through annihilation and production processes. The time
evolution of the number density of the particles is given by
Boltzmann equation and can be evaluated knowing the thermally
averaged total annihilation cross section. The amount of
neutralinos should not be too large to overclose the Universe
and, at the same time, it should be large enough to produce
the right amount of the Dark matter.  This serves as a very
severe bound on SUSY parameters and leaves a very narrow band
in the $m_0 - m_{1/2}$ plane
(WMAP-band)~\cite{Wband1,Wband2,Wband3,Wband4,Wband5,Wband6}.

Having in mind the above mentioned constraints one can find
the most preferable region of the parameter space by
minimizing the $\chi^2$ function. It is remarkable
that all these constraints can be fulfilled
simultaneously.

In what follows, we consider the $m_0 - m_{1/2}$ plane and
find the allowed region in this plane. Each point at this
plane corresponds to a fixed set of parameters and allows one
to calculate the spectrum, cross-sections, etc.

\begin{figure}[b]
\centerline{\psfig{file=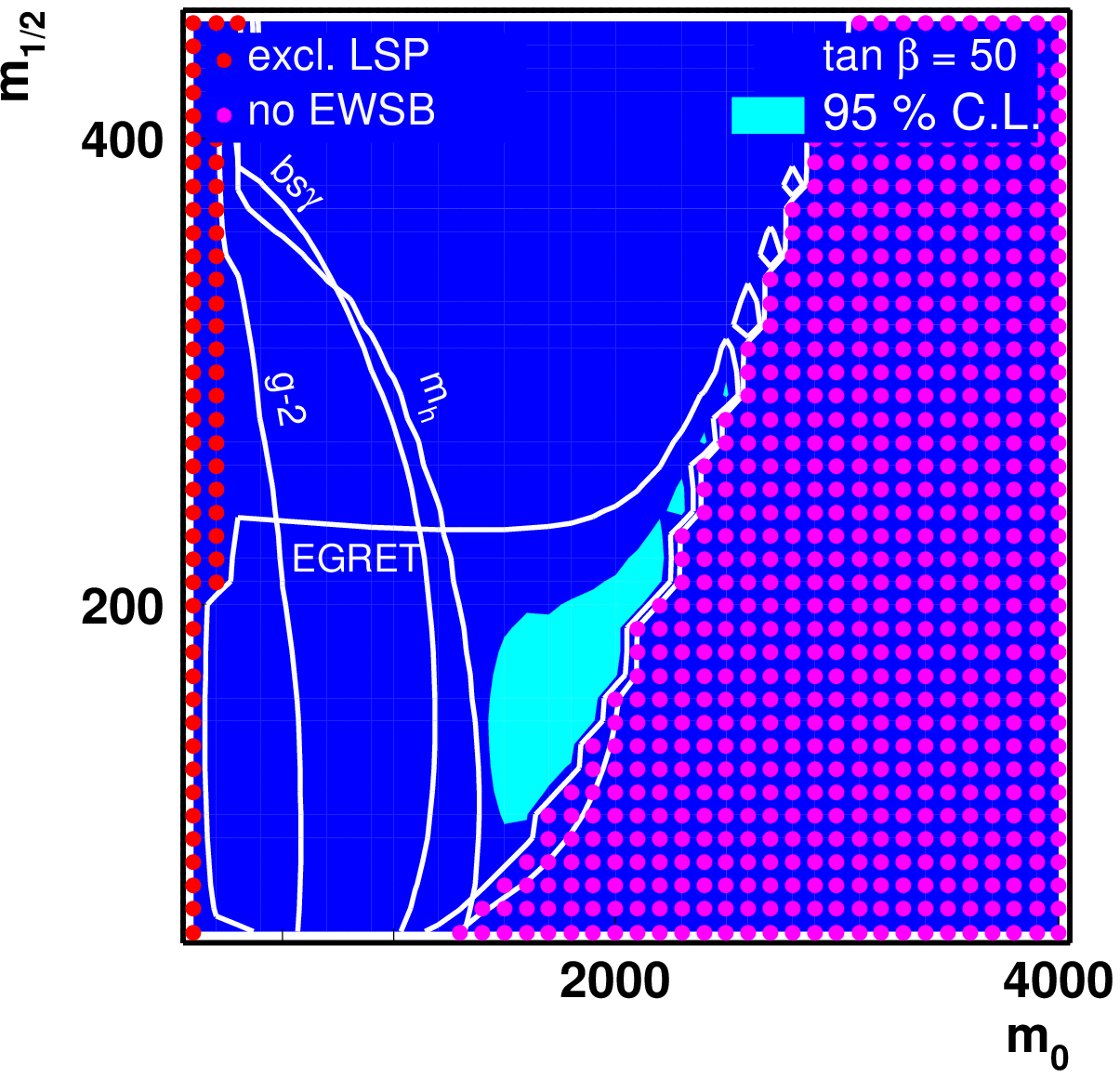,width=0.47\textwidth}
\hspace{0.03\textwidth}
\psfig{file=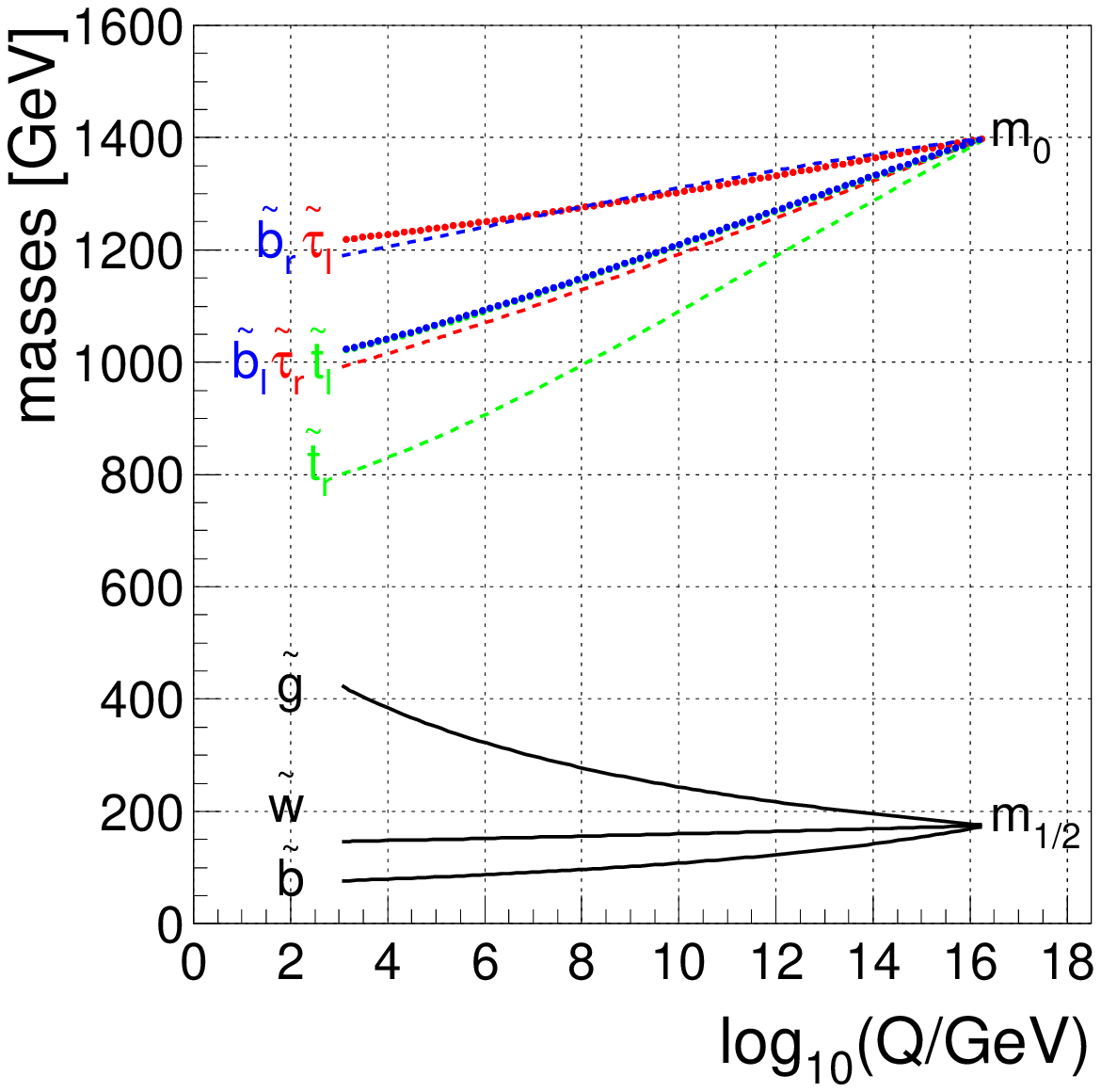,width=0.47\textwidth}}
\vspace*{8pt} \caption{1.MSSM parameter space. The light
shaded (blue) area is the region preferred by EGRET data for
$\tan\beta=50$, $\mu>0$ and $A_0=0$. The excluded regions
where the stau would be the LSP, or EWSB fails, or the Higgs
boson is too light are indicated by the dots. 2. Running
masses of squarks and gauginos with values of $m_1/2$ and
$m_0$ compatible with EGRET data. \label{r}}
\end{figure}

New constraints also comes from astrophysics. According to
recent data from EGRET collaboration on diffuse gamma ray
flux, there is a clear isotropic excess for energies above 1
GeV in comparison with the expectations from conventional
galactic models. If one conjectures that it is
due to the Dark Matter (neutralino) annihilation, then from
the position of the excess  one can fit the preferable
neutralino mass which happens to be around 80 GeV which in
turn restricts the value of $m_{1/2}$. It appears
that the most preferable values are about $m_{1/2}=180$~GeV
and $m_0 \simeq 1400$~GeV (Fig.\ref{r}).

This region has not been intensively studied yet, though it is
within the reach of Tevatron and forthcoming LHC. It appears
to be very interesting phenomenologically, because of the mass
splitting between light gauginos and heavy squarks and
sleptons (see Fig.\ref{r} right). The cross-sections for
chargino and neutralino production in this case are relatively
large not being suppressed by masses and being comparable with
squark and gluino production. The latter being enhanced by
strong interactions remains suppressed by heaviness of
squarks. This means that in the EGRET region leptonic channels
are not suppressed and so might give  clear leptonic signature
for supersymmetry in the upcoming LHC experiments. Below we
study these cross-sections and the main decay modes in detail.

\section{Neutralino and chargino production and decay modes}

Owing to small $m_{1/2}$ and large $m_0$ in the chosen region
of parameter space we deal with  relatively light gauginos and
heavy squarks (Fig.\ref{r}). Charginos and neutralinos are,
respectively, two and four eigenstates of the corresponding
mass matrices. The neutralino mass matrix is

\begin{equation}
M^{(0)}\!=\!\left(
\begin{array}{cccc}
M_1 & 0 & -M_{\!Z} c_\beta s_W & M_{\!Z} s_\beta s_W \\
0 &
M_2 & M_{\!Z} c_\beta c_W   & -M_{\!Z} s_\beta c_W  \\
-M_{\!Z} c_\beta s_W & M_{\!Z} c_\beta c_W  & 0 & -\mu \\
M_{\!Z} s_\beta s_W & -M_{\!Z} s_\beta c_W  & -\mu & 0 \\
\end{array} \right)\!,
\label{fff}
\end{equation}
where $c_\beta=\cos\beta$, $s_\beta=\sin\beta$. $s_W$ and
$c_W$ are the sine and cosine of the Weinberg angle
respectively. Four eigenstates of this matrix are the four
types of physical particles. The lightest one
$\tilde\chi_1^0$,  is the LSP. In the chosen region of small
$m_{1/2}$ the lightest neutralino mass is around $70$~GeV
which perfectly fits the EGRET data. For
$m_{1/2}=180$~GeV the values of gaugino masses are
$M_1=120$~GeV and $M_2~=~250$~GeV.

The lightest neutralino in our case is mostly bino, the second
neutralino is mostly wino, while the third and the fourth
are mostly higgsinos (see Table \ref{a}).

\begin{table}[b]
\tbl{The bino, wino, first and second higgsino fractions
of neutralinos.}
{\begin{tabular}{@{}cccccc@{}} \toprule
 & \hphantom{aaa}Bino\hphantom{aaa} & \hphantom{aaa}Wino\hphantom{aaa} & 1$^{\rm st}$ higgsino & 2$^{\rm nd}$ higgsino & mass (GeV)\\
\colrule
$\chi^0_1$ & $90\%$            & $\hphantom{0}1\%$ & $\hphantom{0}8\%$ & $\hphantom{0}1\%$ & \hphantom{0}70 \\
$\chi^0_2$ & $\hphantom{0}5\%$ & $71\%$            & $18\%$            & $\hphantom{0}6\%$ & 125 \\
$\chi^0_3$ & $\hphantom{0}1\%$ & $\hphantom{0}2\%$ & $46\%$            & $51\%$            & 220 \\
$\chi^0_4$ & $\hphantom{0}3\%$ & $25\%$            & $30\%$            & $42\%$            & 250 \\
\botrule
\end{tabular} \label{a}}
\end{table}

\begin{figure}[t]
\centerline{\psfig{file=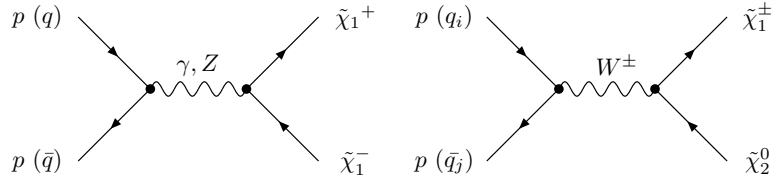,width=0.8\textwidth}}
\vspace*{8pt}
\caption{Leading chargino and neutralino production processes. \label{prod}}
\end{figure}

Two egienstates of the chargino mass matrix
\begin{equation}
M^{(c)}\!=\!\left(
\begin{array}{cc}
M_2 & \sqrt{2}M_W s_\beta \\ \sqrt{2}M_W c_\beta & \mu
\end{array} \right).
\label{ccm}
\end{equation}
are the two charginos $\tilde\chi^\pm_{1,2}$. Their masses are
given by the following formula:
\begin{eqnarray}
&& M^2_{1,2}\!=\!\frac{1}{2}\Big[M^2_2\!+\!\mu^2\!+\!2M^2_{\!W} \mp\\
&& \left.\! \sqrt{(M^2_2\!\!-\!\!\mu^2)^2\!\!+\!4M^4_{\!W} \cos^2{\! 2\beta}
\!\!+\!4M^2_{\!W}(M^2_2\!\!+\!\mu^2\!\!+\!2M_2\mu \sin{\! 2\beta})}\right]
\nonumber
\end{eqnarray}
The first chargino is relatively light due to the gaugino
component.

To calculate the cross-sections for sparticle production at
hadron colliders we use the CalcHEP~2.3.7.
package~\cite{calchep} which takes into account parton
distributions inside protons~\cite{MRST}. First we selected
three leading types of chargino and neutralino production
processes (Fig.~\ref{prod}).

\begin{figure}[b]
\centerline{\psfig{file=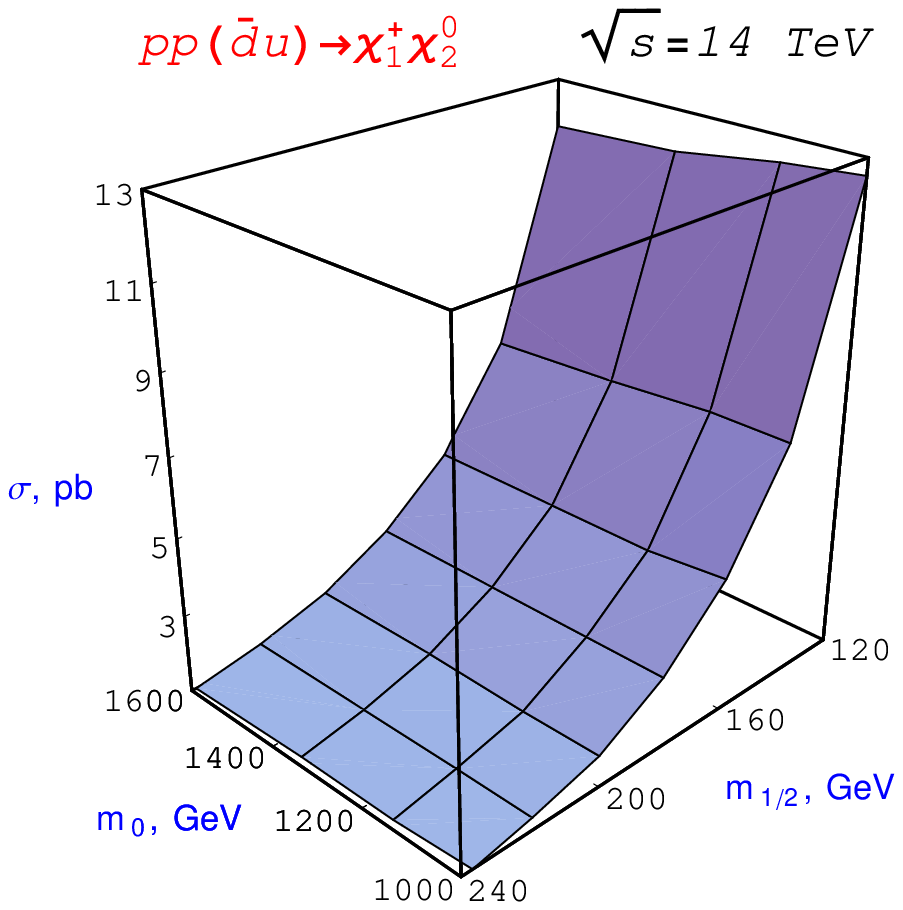,width=0.47\textwidth}
\hspace{0.03\textwidth}
\psfig{file=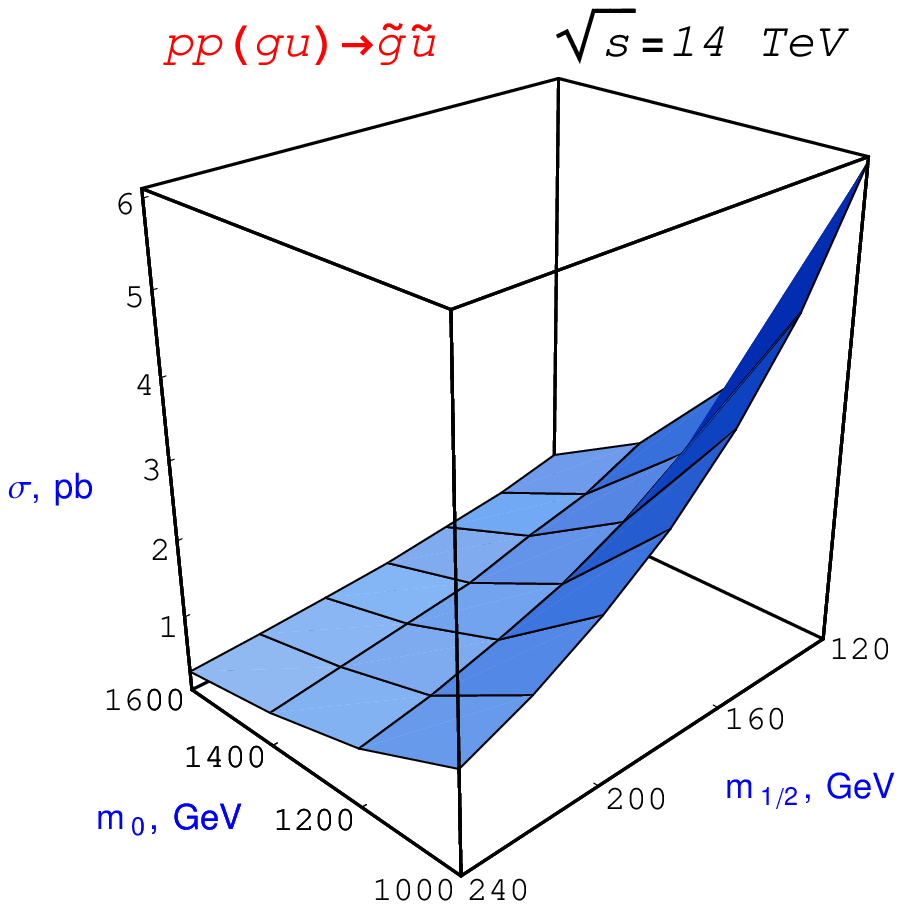,width=0.47\textwidth}}
\vspace*{8pt}
\caption{The cross-section for chargino and neutralino production at LHC for
$\sqrt{s}=14$ TeV as functions of $m_{1/2}$ and  $m_{0}$ for $\tan\beta=51$,
$A_0=0$ and sign($\mu$)=1. \label{b}}
\end{figure}

\begin{figure}[tb]
\centerline{\psfig{file=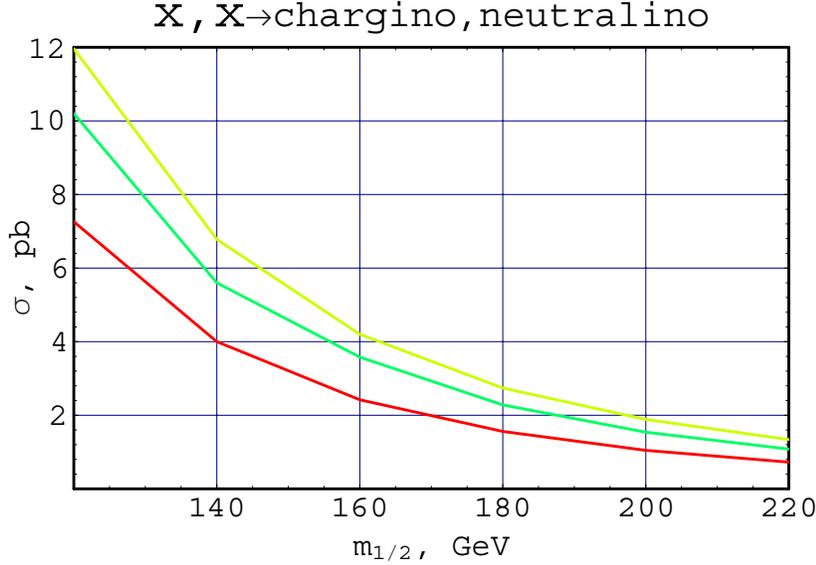,width=0.85\textwidth}}
\vspace*{8pt}
\caption{The cross-section dependence on $m_{1/2}$ for chargino and neutralino
production. The upper yellow line is for $pp(q\bar{q})\rightarrow
\tilde{\chi}_{2}^{0}\tilde{\chi}_{1}^{+}+X$, the red line in the bottom is for
$pp(q\bar{q})\rightarrow \tilde{\chi}_{2}^{0}\tilde{\chi}_{1}^{-}+X$, and the green line in the
middle is for $pp(q\bar{q})\rightarrow\tilde{\chi}_{1}^{+}\tilde{\chi}_{1}^{-}+X$
in case of $m_{0}=1400$~GeV, $\tan\beta=51$, $A_0=0$ and sign($\mu$)=1. \label{c}}
\end{figure}

The calculations are made for proton-proton collisions at the
center of mass energy of $\sqrt{s}=14$~TeV (LHC).  As one can
see from Fig.~\ref{b}, the cross sections for neutralino and
chargino production processes slightly depend on $m_0$ and
strongly depend on $m_{1/2}$; thus, for the future reference
we fix the value of $m_0=1400$~GeV. The cross section
dependence on $m_{1/2}$ for chargino and neutralino production
processes is shown in Fig.\ref{c}.

\begin{table}[b]
\tbl{The 1$^{\rm st}$ chargino and the 2$^{\rm nd}$ neutralino decay modes and partial widths for
$m_{0}=1400$~GeV, $m_{1/2}=180$~GeV, $\tan\beta=51$, $A_0=0$ and sign($\mu$)=1.}
{\begin{tabular}{@{}cccc@{}} \toprule
Initial particle & Decay mode & Branching ratio & Partial width (GeV)\\
\colrule
$\chi^\pm_1$ &
$\begin{array}{l}
\chi^0_1 \ \bar q_i \ q_k \\ \chi^{1}_0 \ \ell \ \nu
\end{array}$ &
$\begin{array}{l}
67\% \\ 33\%
\end{array}$ &
$\begin{array}{l}
0.35\times10^{-4} \\ 0.17\times10^{-4}
\end{array}$ \\
\colrule
$\chi^0_2$ &
$\begin{array}{l}
\chi^0_1 \ q \ \bar q  \\ \chi^0_1 \ \bar\ell \ \ell  \\ \chi^0_1
\ \bar\nu \ \nu
\end{array}$ &
$\begin{array}{l}
70\% \\ 10\% \\ 20\%
\end{array}$ &
$\begin{array}{l}
0.66\times10^{-5} \\ 0.10\times10^{-5} \\ 0.20\times10^{-5}
\end{array}$ \\
\botrule
\end{tabular} \label{h}}
\end{table}

\begin{table}[th]
\tbl{Production of the 1$^{\rm st}$ chargino and the 2$^{\rm nd}$
neutralino with subsequent cascade decays.}
{\begin{tabular}{@{}cc@{}}
\toprule
Process & Final states \\
\colrule
\psfig{file=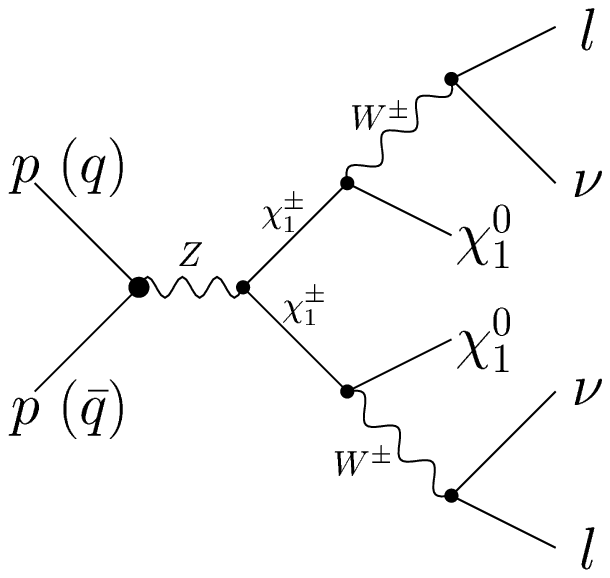,width=38mm,height=30mm} &
\raisebox{15mm}{
$\begin{array}{c}
2\ell \\ 2\nu \\ \Big/\hspace{-0.3cm E_T} \\ \\ \sigma \approx 0.25~{\rm pb}
\end{array}$} \\
\colrule
\psfig{file=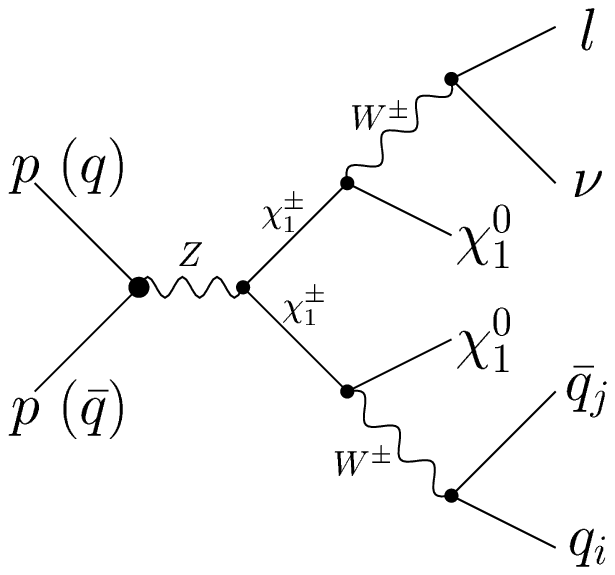,width=38mm,height=30mm} &
\raisebox{15mm}{
$\begin{array}{c}
\ell \\ \nu \\ 2j \\ \Big/\hspace{-0.3cm E_T} \\ \\ \sigma \approx 0.50~{\rm pb}
\end{array}$} \\
\colrule
\psfig{file=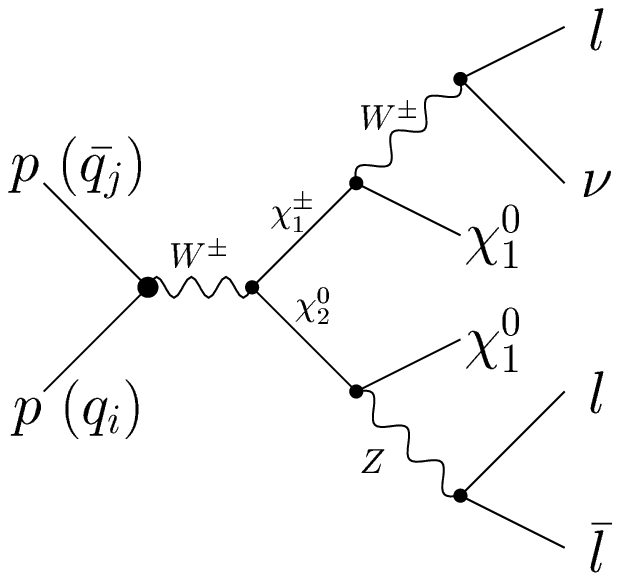,width=38mm,height=30mm} &
\raisebox{15mm}{
$\begin{array}{c}
3\ell \\ \nu \\ \Big/\hspace{-0.3cm E_T} \\ \\ \sigma \approx 0.14~{\rm pb}
\end{array}$} \\
\botrule
\end{tabular}
\hspace{2mm}
\begin{tabular}{@{}cc@{}}
\toprule
Process & Final states \\
\colrule
\psfig{file=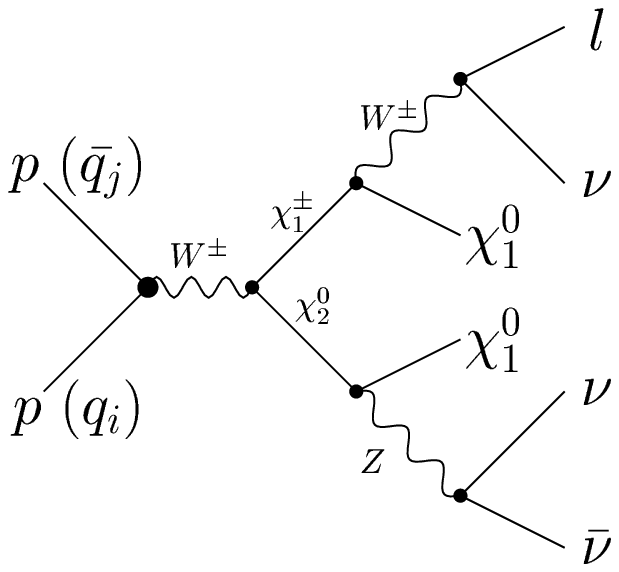,width=38mm,height=30mm} &
\raisebox{15mm}{
$\begin{array}{c}
\ell \\ 3\nu \\ \Big/\hspace{-0.3cm E_T} \\ \\ \sigma \approx 0.28~{\rm pb}
\end{array}$} \\
\colrule
\psfig{file=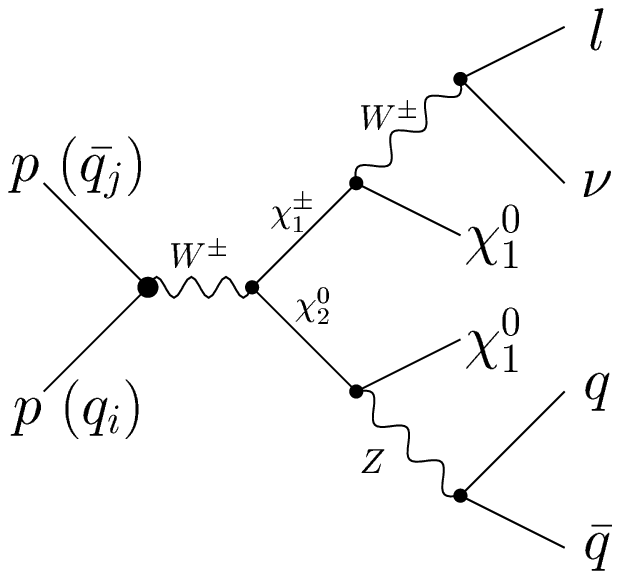,width=38mm,height=30mm} &
\raisebox{15mm}{
$\begin{array}{c}
\ell \\ \nu \\ 2j \\ \Big/\hspace{-0.3cm E_T} \\ \\ \sigma \approx 1.0~{\rm pb}
\end{array}$} \\
\colrule
\psfig{file=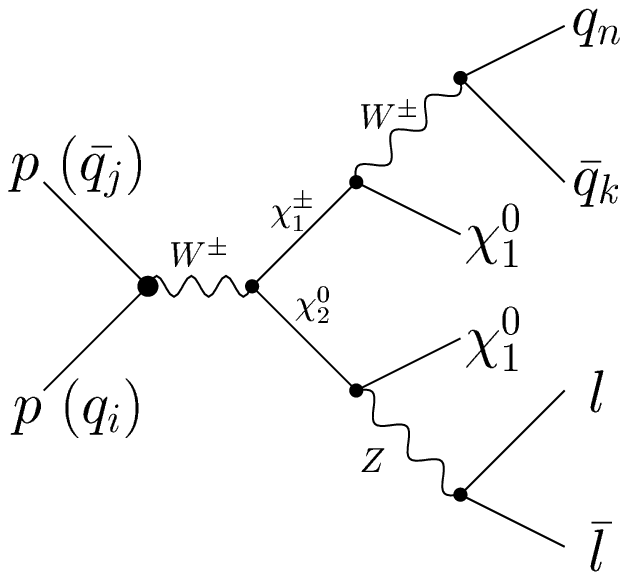,width=38mm,height=30mm} &
\raisebox{15mm}{
$\begin{array}{c}
2\ell \\ 2j \\ \Big/\hspace{-0.3cm E_T} \\ \\ \sigma \approx 0.29~{\rm pb}
\end{array}$} \\
\botrule
\end{tabular}
\label{j}}
\end{table}

In this region of parameter space the cross section for
production of the second neutralino alongside with the first
chargino is larger, owing to the mixing parameters, than that
when the lightest neutralino is produced. After being produced
the light chargino and the second neutralino rapidly decay
into the Standard Model particles and the LSP. The
decay modes and partial widths obtained from ISAJET~\cite{isajet}
are shown in Table~\ref{h}.

\begin{figure}[tb]
\centerline{\psfig{file=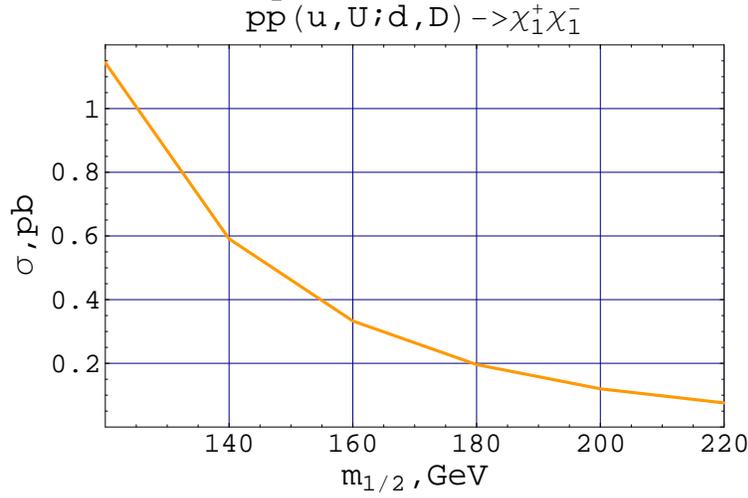,width=0.85\textwidth}}
\vspace*{8pt}
\caption{Cross section dependence on $m_{1/2}$
for chargino pair production $p\bar{p} \rightarrow\chi_{1}^{+}\chi_{1}^{-}$ at TeVatron
for $m_0$ = 1400 GeV,  $\tan\beta=51$, $A_0=0$ and sign($\mu$)=1. \label{Dd}}
\end{figure}

Chargino mostly decays into LSP and via virtual $W$ boson into either
quarks of different flavor or  a neutrino--lepton pair. The second neutralino decays into
LSP and via virtual $Z$-boson into either quark--antiquark pair or lepton--antilepton
pair. Different types of chargino and neutralino cascade decays are shown in
Table~\ref{j}.

Thus, in the chosen region of parameter space due to the
relatively large neutralino and chargino production cross
sections and branching ratios to leptons and LSP, one can have
an unexpectedly large number of outgoing leptons. Moreover,
some of the pure leptonic events, like one or three charged
leptons clearly differ from the Standard Model background
(pair $W$ and $Z$ production) by \textit{large} missing energy,
thus being a convincing indication to the new physics.

The same analysis was conducted for the pro\-ton-anti\-pro\-ton
collisions at the Tevatron
center of mass energy $\sqrt{s}=2$~TeV. The Tevatron energy is
not sufficient to produce heavy
squarks or sleptons but is sufficient to produce light
char\-gi\-nos and neutralinos. The
cross-section dependence for chargino pair production
on $m_{1/2}$ is shown in
Fig.~\ref{Dd}. As one can see the cross-section is
around 0.1--1~pb. The chargino has the
same decay modes as discussed above but ten times smaller
cross section compared
to LHC. However, the present Tevatron luminosity is not
enough to extract the signal from
the background, and one can get only lower limits on chargino masses
(117 GeV at the 95\% C.L.)~\cite{D0}.

\section{Squark and gluino production and decay modes}

Squarks of the 3rd generation are two eigenstates of the mass
matrices
\begin{eqnarray}
&&\left(
\begin{array}{cc}
\tilde m_{tL}^2& m_t(A_t-\mu\cot \beta )\\
m_t(A_t-\mu\cot \beta ) & \tilde m_{tR}^2
\end{array}
\right), \\
&&\left(
\begin{array}{cc} \tilde  m_{bL}^2& m_b(A_b-\mu\tan \beta )\\
m_b(A_b-\mu\tan \beta ) & \tilde  m_{bR}^2
\end{array}
\right),
\end{eqnarray}
where
\begin{eqnarray}
  \tilde m_{tL}^2&=&\tilde{m}_Q^2+m_t^2+\frac{1}{6}(4M_W^2-M_Z^2)\cos
  2\beta , \nonumber \\
  \tilde m_{tR}^2&=&\tilde{m}_U^2+m_t^2-\frac{2}{3}(M_W^2-M_Z^2)\cos
  2\beta ,\\
  \tilde m_{bL}^2&=&\tilde{m}_Q^2+m_b^2-\frac{1}{6}(2M_W^2+M_Z^2)\cos
  2\beta , \nonumber \\
  \tilde m_{bR}^2&=&\tilde{m}_D^2+m_b^2+\frac{1}{3}(M_W^2-M_Z^2)\cos
  2\beta . \nonumber
\label{sq}
\end{eqnarray}
The first terms here are the soft SUSY breaking parameters,
the second terms are the ordinary quark masses and the third
ones are the so called $D$-terms. Similar formulae can
be written for the first and second generations.

The nonvanishing Yukawa couplings result in the large mixing
and splitting of mass eigenstates for the third generation of
squarks. As a result one of the eigenstates becomes lighter
than the others. This leads to a remarkable consequence: the
decay branching ratios for the third generation of squarks are
several times bigger than those for the first two generations
as we will discuss below.

\begin{figure}[ptb]
\centerline{\psfig{file=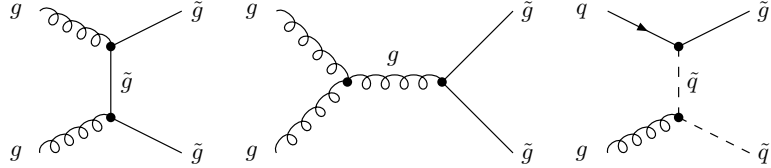,width=0.8\textwidth}}
\vspace*{8pt}
\caption{Leading types of strongly interacting sparticles
production processes. \label{prodq}}
\end{figure}

The cross-sections for squark production via proton collisions
at the center of mass energy of $\sqrt{s}=14$~TeV are strongly
suppressed due to the large squark masses. At the same time,
the cross-sections for squark-gluino and pair gluino
production are not that strongly suppressed due to the
relatively light gluino. The leading procceses for production
of strongly interacting spacticles are shown in
Fig.~\ref{prodq}.

\begin{figure}[ptb]
\centerline{\psfig{file=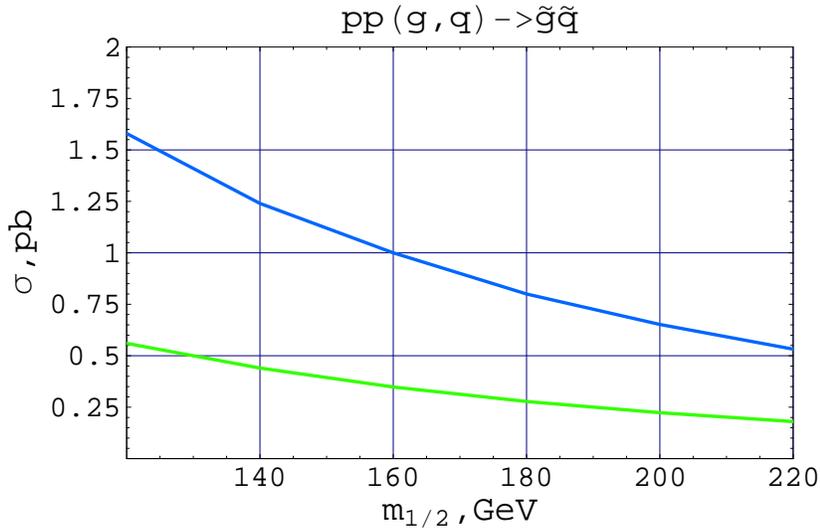,width=0.9\textwidth}}
\vspace*{8pt}
\caption{Cross-section for squark production
as a function of $m_{1/2}$. The upper blue line is for
$pp\ (q\bar{q})\rightarrow
\tilde{g}\tilde{u}+X $, the purple line in the bottom is for
$pp\ (q\bar{q})\rightarrow
\tilde{g}\tilde{d}+X$ in case of $m_0$=1400~GeV $\tan\beta=51$,
$A_0=0$ and sign($\mu$)=1. \label{e}}
\end{figure}

\begin{figure}[ptb]
\centerline{\psfig{file=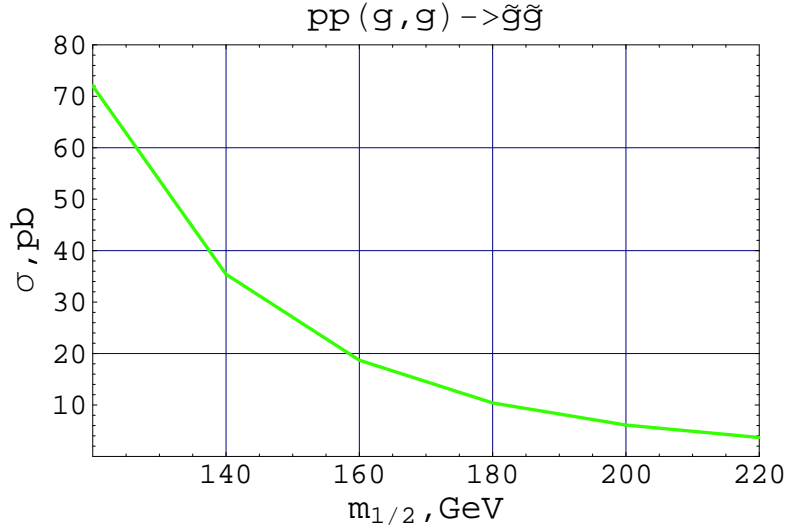,width=0.9\textwidth}}
\vspace*{8pt}
\caption{Cross-section  for gluino production $pp\ (gg)\rightarrow
\tilde{g}\tilde{g}+X $ as a function of $m_{1/2}$
in case of $m_0$=1400~GeV $\tan\beta=51$, $A_0=0$ and sign($\mu$)=1. \label{f}}
\end{figure}

The cross-sections for squark and gluino production depend on
$m_{1/2}$ much stronger than on $m_0$. As it was mentioned
above we fixed the value of $m_0~=~1400$~GeV. The dependence of
squark and gluino production cross-section on $m_{1/2}$ is
shown in Fig.~\ref{e}.

One can see that the cross-section for squark
and gluino production is of the same order
of magnitude as the cross section for chargino
and neutralino production. However,
further squarks and gluinos are the main source
of outgoing jets in sparticle decay while
charginos and second neutralinos are the main
source of leptons.

At the same time, in the small $m_{1/2}$ region
light gluino  production cross-section
in proton collisions via gluon fusion
(Fig.~\ref{f}) is by almost two orders of magnitude
larger than all other sparticle production cross-sections.

After being produced gluinos rapidly decay.
Different types of gluino decay modes
 obtained from ISAJET~\cite{isajet}
are shown in Table~\ref{g}.  There are two leading decay
modes. First of them is when gluino decays to bottom quark and
via virtual sbottom into the second neutralino and anti-bottom
quark, while in the second case it decays into quark and via
virtual squark to the first chargino and anti-quark of a
different flavor. Charginos and neutralinos produced in gluino
decay processes may decay not only to quarks but also to
leptons. Hence, gluino cascade decays might have a different
number of leptons and jets in the final states, as shown in
Table~\ref{n}. This slightly differs chargino and neutralino
decays which have pure leptonic final states.

\begin{table}[b]
\tbl{Gluino and top-quark decay modes and partial widths.}
{\begin{tabular}{@{}cccc@{}} \toprule
Initial particle & Decay mode & Branching ratio & Partial width (GeV)\\
\colrule
$\tilde g$ &
$\begin{array}{l}
\chi^0_2 \ \bar b \ b \\
\chi^0_2 \ \bar q \ q \\
\chi^\pm_1 \ \bar q_i \ q_k \\
\chi^\pm_1 \ \bar t \ b
\end{array}$ &
$\begin{array}{l}
16\% \\ 24\% \\ 30\% \\ 6.6\%
\end{array}$ &
$\begin{array}{l}
0.14\times10^{-4} \\ 0.20\times10^{-4} \\ 0.27\times10^{-4} \\ 0.55\times10^{-4}
\end{array}$ \\
\colrule
$t$ &
$\begin{array}{l}
b \ \bar q_i \ q_k \\ b \ \ell \ \nu
\end{array}$ &
$\begin{array}{l}
67\% \\ 33\%
\end{array}$ &
$\begin{array}{l}
0.8 \\ 0.4
\end{array}$ \\
\botrule
\end{tabular} \label{g}}
\end{table}

\begin{table}[th]
\tbl{Cascade decays for pair of gluinos.}
{\begin{tabular}{@{}cc@{}}
\toprule
Process & Final states \\
\colrule
\psfig{file=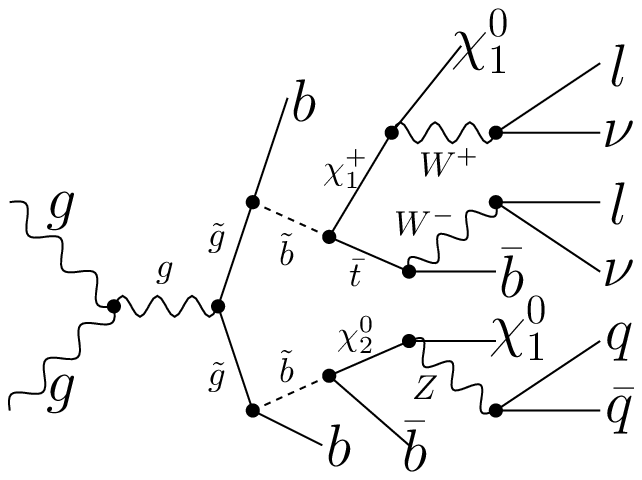,width=38mm,height=30mm} &
\raisebox{15mm}{
$\begin{array}{c}
2\ell \\ 2\nu \\ 6j \\ \Big/\hspace{-0.3cm E_T} \\ \\ \sigma \approx 0.008~{\rm pb}
\end{array}$} \\
\colrule
\psfig{file=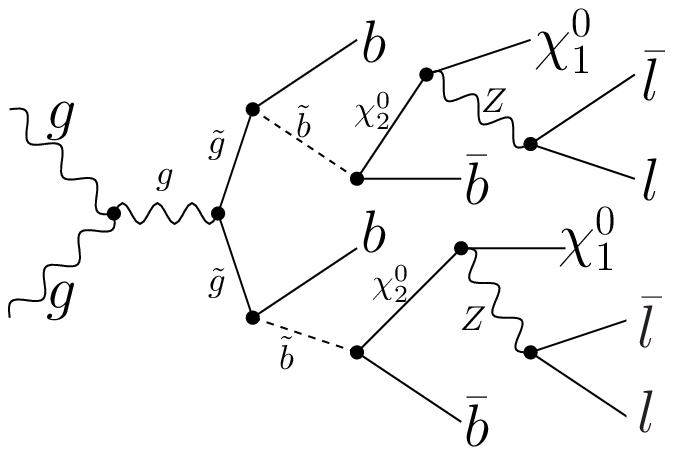,width=38mm,height=30mm} &
\raisebox{15mm}{
$\begin{array}{c}
4\ell \\ 4j \\ \Big/ \hspace{-0.3cm E_T} \\ \\ \sigma \approx 0.003~{\rm pb}
\end{array}$} \\
\colrule
\psfig{file=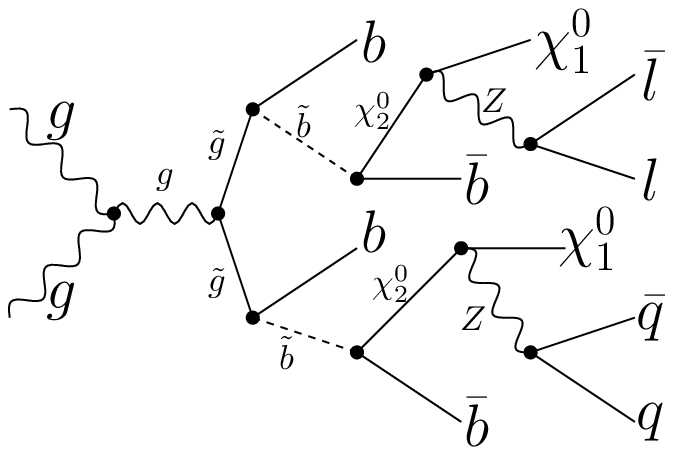,width=38mm,height=30mm} &
\raisebox{15mm}{
$\begin{array}{c}
2\ell \\ 6j \\ \Big/ \hspace{-0.3cm E_T} \\ \\ \sigma \approx 0.019~{\rm pb}
\end{array}$} \\
\botrule
\end{tabular}
\hspace{2mm}
\begin{tabular}{@{}cc@{}}
\toprule
Process & Final states \\
\colrule
\psfig{file=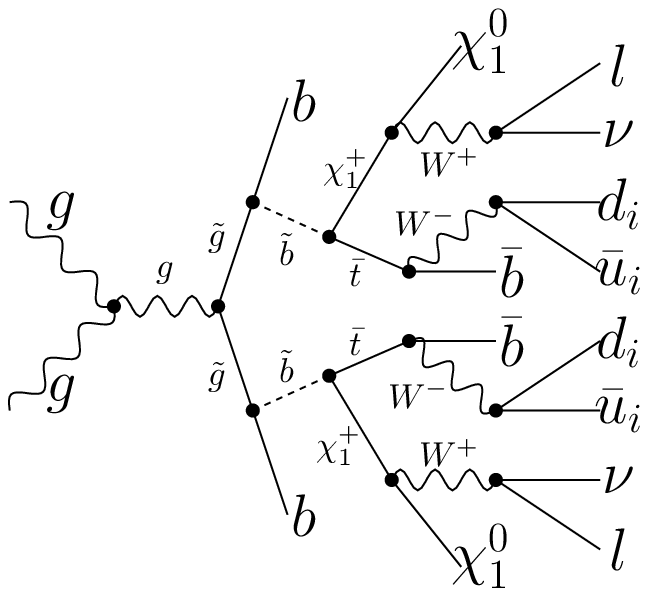,width=38mm,height=30mm} &
\raisebox{15mm}{
$\begin{array}{c}
2\ell \\ 2\nu \\ 8j \\ \Big/\hspace{-0.3cm E_T} \\ \\ \sigma \approx 0.002~{\rm pb}
\end{array}$} \\
\colrule
\psfig{file=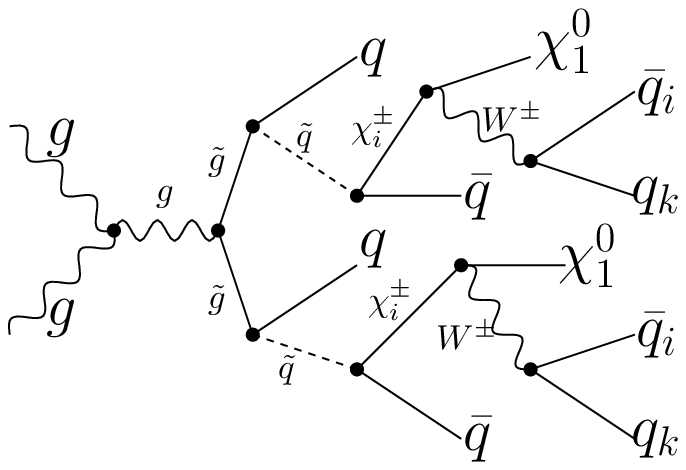,width=38mm,height=30mm} &
\raisebox{15mm}{
$\begin{array}{c}
8j \\ \Big/ \hspace{-0.3cm E_T} \\ \\ \sigma \approx 0.40~{\rm pb}
\end{array}$} \\
\colrule
\psfig{file=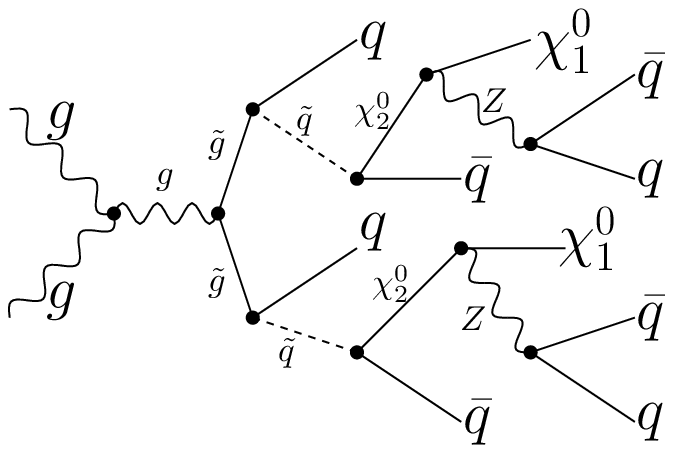,width=38mm,height=30mm} &
\raisebox{15mm}{
$\begin{array}{c}
8j \\ \Big/ \hspace{-0.3cm E_T} \\ \\ \sigma \approx 0.29~{\rm pb}
\end{array}$} \\
\botrule
\end{tabular}
\label{n}}
\end{table}

\begin{figure}[tb]
\centerline{\psfig{file=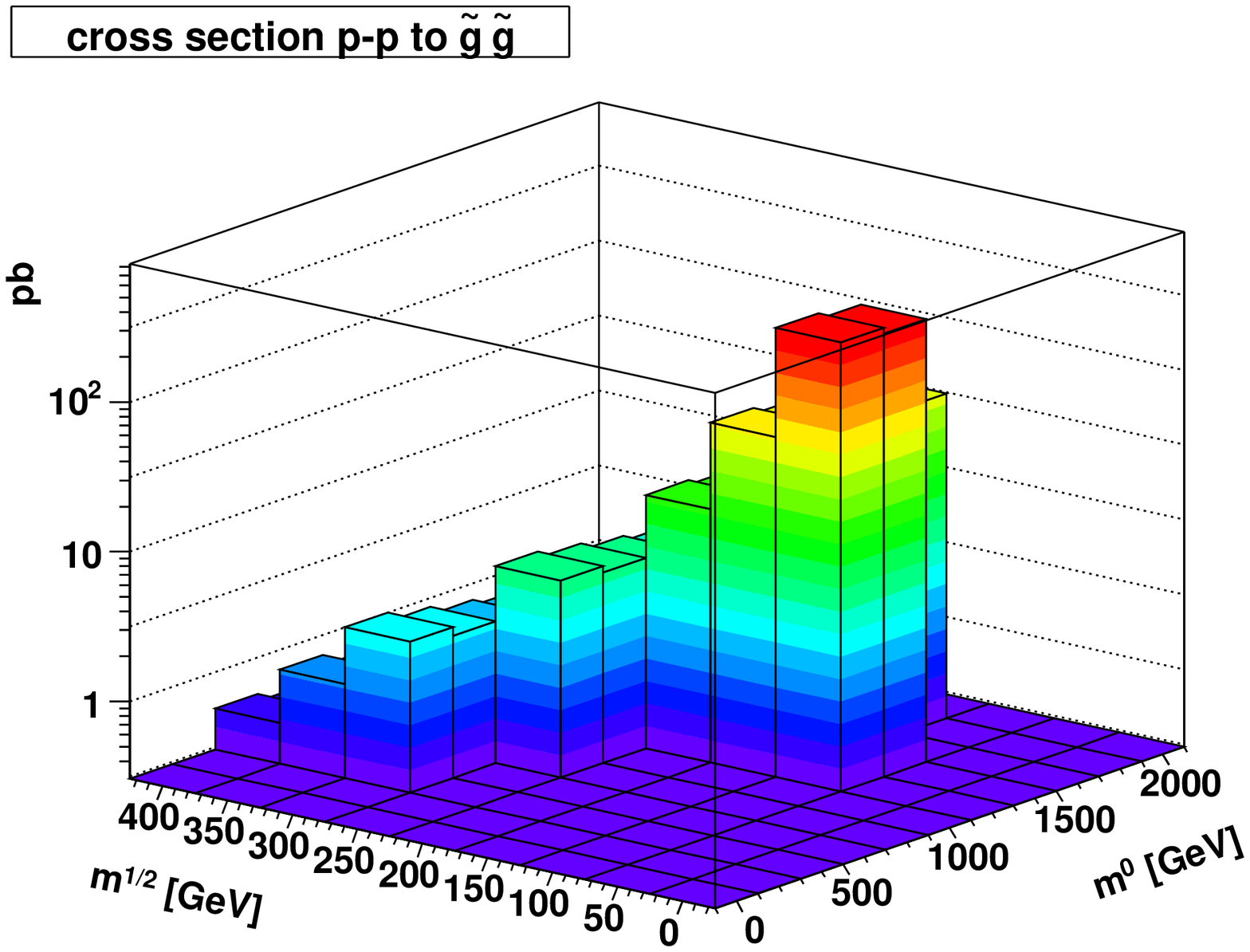,width=0.45\textwidth}
\hspace{0.05\textwidth}
\psfig{file=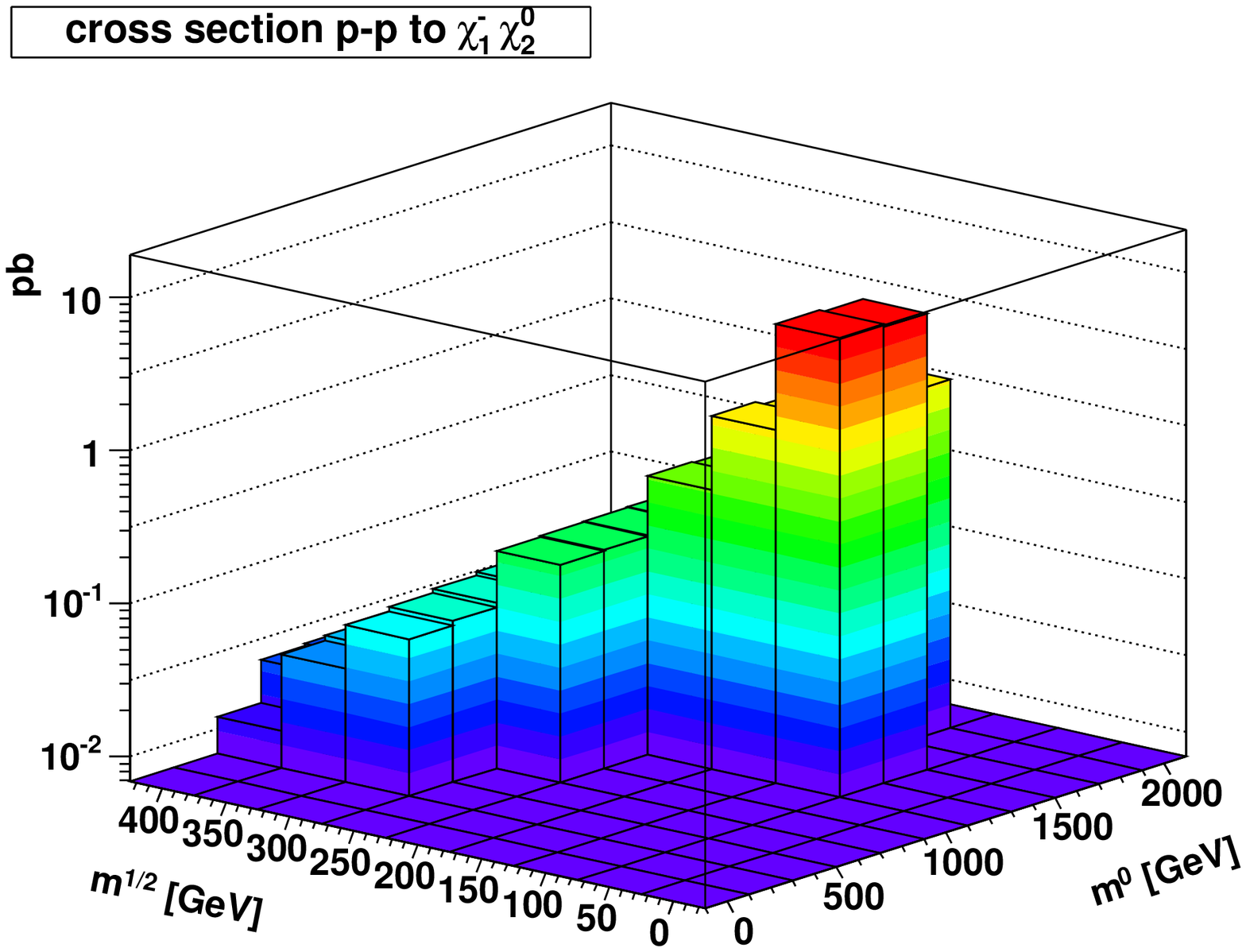,width=0.45\textwidth}}
\centerline{\psfig{file=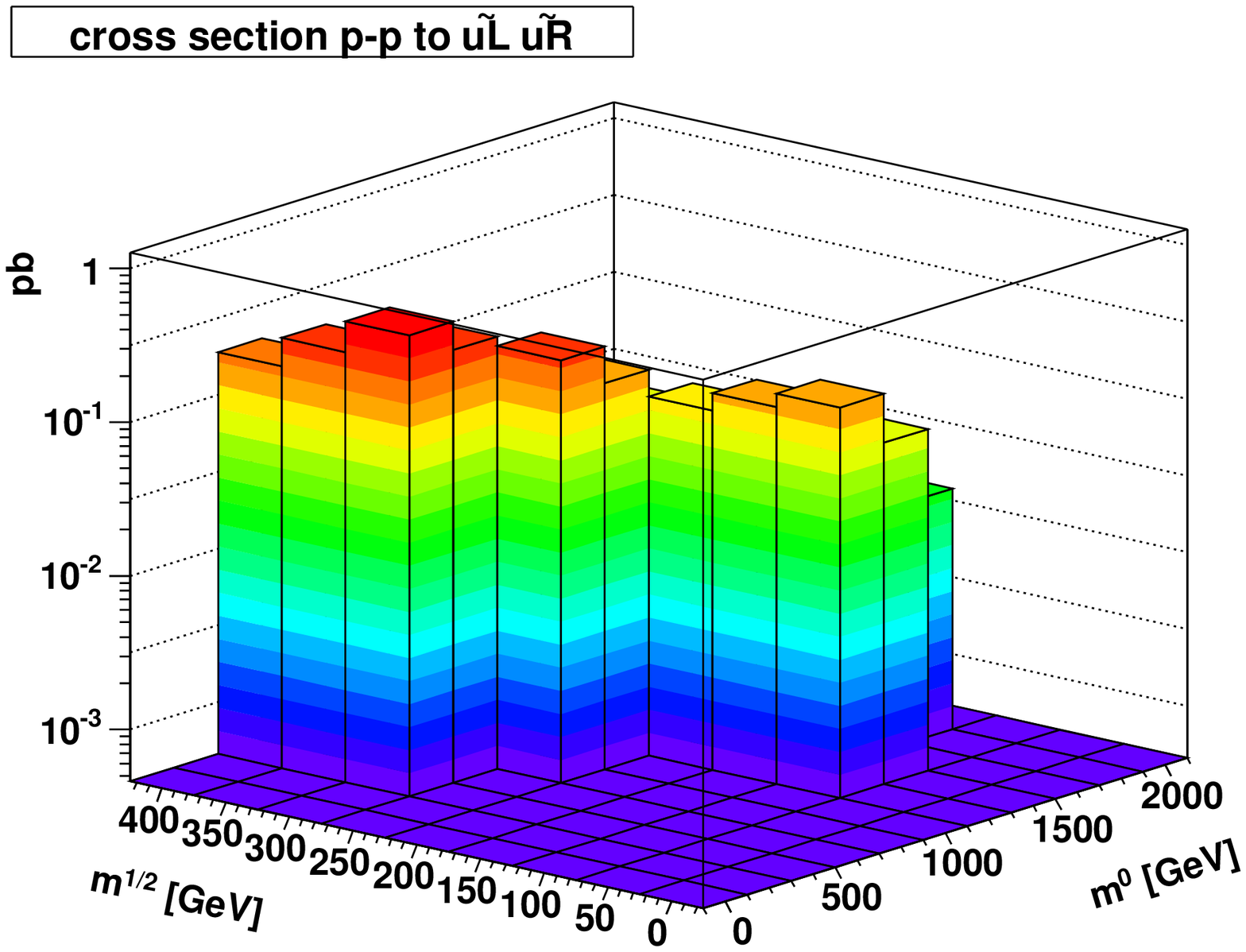,width=0.45\textwidth}
\hspace{0.05\textwidth}
\psfig{file=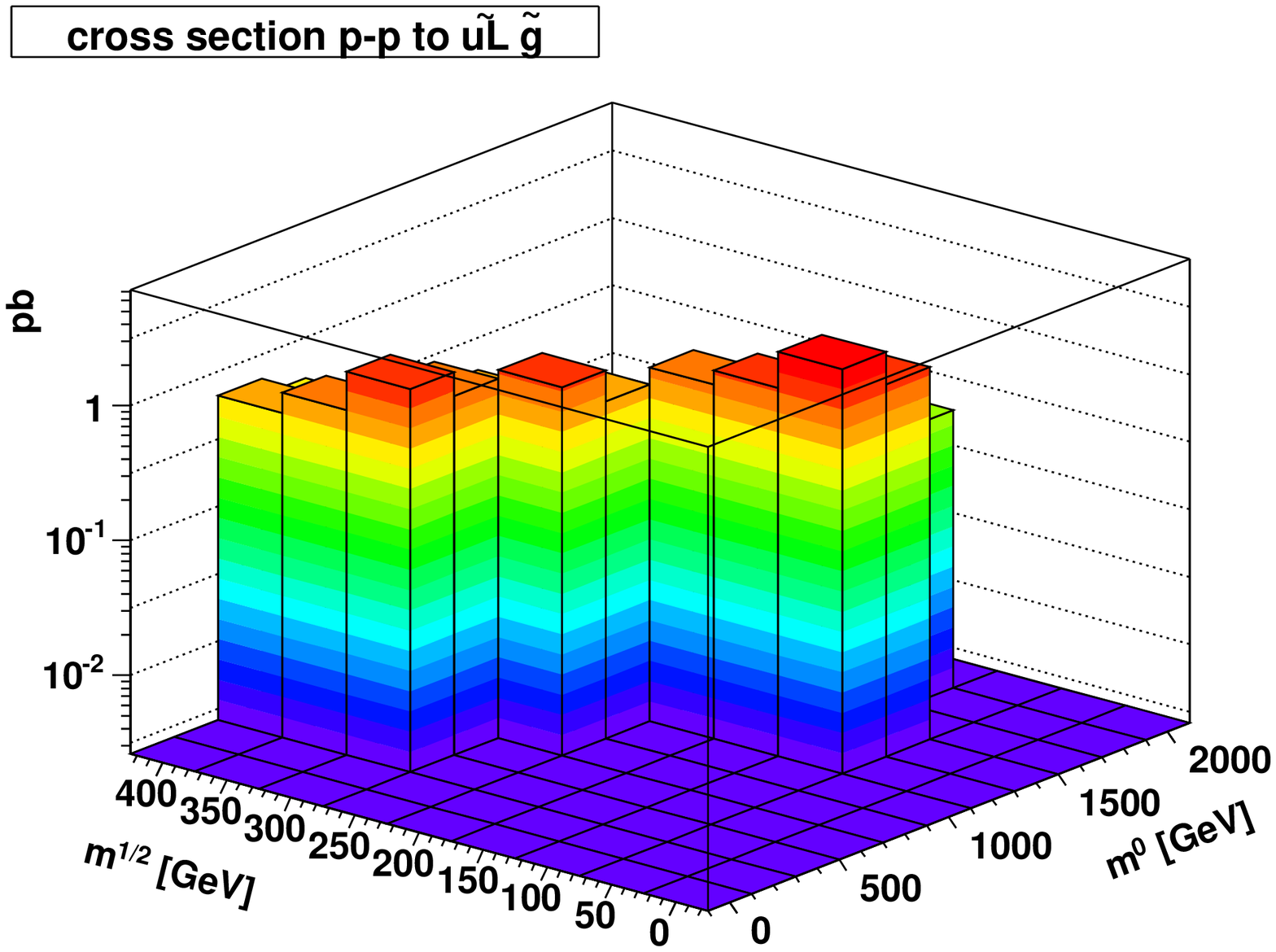,width=0.45\textwidth}}
\vspace*{8pt}
\caption{Cross-section dependence on $m_{1/2}$, $m_{0}$ for sparticle production
processes for  $\tan\beta=51$, $A_0=0$ and sign($\mu$)=1. \label{Mss}}
\end{figure}

As one can see from Fig.~\ref{Mss}, the cross-sections for all
sparticle production processes drop drastically with increase
of $m_{1/2}$. This is especially true for the production
pro\-ces\-ses of weakly interacting sparticles: charginos and
second neutralinos due to the large gaugino fraction.
Alongside with them drops the number of leptonic events above
the Standard Model background. With
increase of $m_{1/2}$ gluino production cross-section also
becomes suppressed by heavy gluino mass. Most of the final
state cross-sections in gluino cascade decays with a variety
of leptons and jets in the final states also drop by $2-3$
orders of magnitude with the increase of $m_{1/2}$ up to $400$~GeV.

Thus, it looks like only a small $m_{1/2}$  region
might bring a relatively large number
of leptoninc events coming above the SM background.
Moreover, from the analysis conducted
for the Tevatron (see Fig.~\ref{Dd}) in this region
of parameter space supersymmetry is
already within a reach of the modern hadron colliders
and the only reason why it is not
seen is the low luminosity of the Tevatron.

Remarkable that in jet production one mainly has $b$-jets with the branching ratio that
may reach 30 \%. This fact finds its explanation in kinematic factors due to the fact
that $b$-squarks are lighter than those of the first two generations.

\begin{figure}[b]
\centerline{\psfig{file=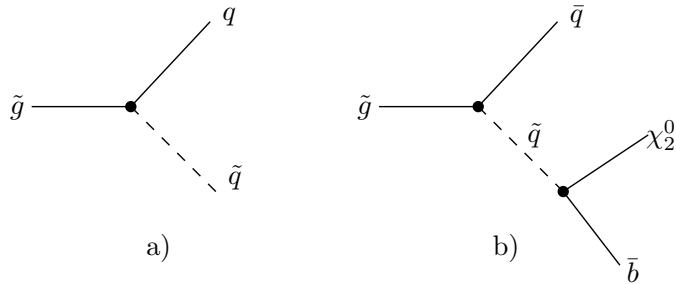,width=0.7\textwidth}}
\vspace*{8pt}
\caption{Leading modes of gluino decay. \label{pic1}}
\end{figure}

Indeed, consider  the process of gluino decay into quarks and
squarks. There are two possibilities: either gluino is heavier
than squarks and can decay on the mass shell (see
Fig.\ref{pic1}a) or it is lighter than squarks and decays into
virtual squark which in turn decays into the second neutralino
and the quark (see Fig.\ref{pic1}b).

In the first case, the width is easily calculable and has the form
\begin{eqnarray}\label{width}
\Gamma(\tilde{g}\rightarrow\tilde{q }q)&\sim&
\frac{1}{m_{\tilde{g}}^3}\sqrt{(m_{\tilde{g}}^2
\!-\!(m_{q}\!-\!m_{\tilde{q}})^2)
(m_{\tilde{g}}^2\!-\!(m_{q}\!+\!m_{\tilde{q}})^2)}
\nonumber \\
&\times&(m_{\tilde{q}}^2-m_{\tilde{g}}^2-m_{q}^2
\pm2\sin{2\theta_{q}}m_{\tilde{g}}m_{q}),
\end{eqnarray}
where $\theta_{q}$ is the squark mixing angle proportional
to the quark mass according to
eq.(\ref{sq}) and $\pm$~sign refers to the first and the
second squark, respectively. The
square root in this equation essentially comes from the phase space.

Since the quark mass is much smaller than the squark
one and can be ignored (except for
the t quark), eq.(\ref{width}) can be reduced to
\begin{equation}
\label{width2}
\Gamma(\tilde{g}\rightarrow\tilde{q }q)
\sim (m_{\tilde{g}}^2-m_{\tilde{q}}^2)^2,
\end{equation}
which clearly shows that the lighter the squark, the bigger
the width. Since the third generation of squarks is lighter
than the first two (mainly due to the RG running induced by
Yukawa couplings and the mixing in squark mass matrix), the
width for the third generation is larger. Besides, the top
quark production is suppressed compared to the bottom one due
to the large top mass which decreases the phase space in
eq.(\ref{width}). So the $b$-quark production is enhanced
compared to other flavours.

The situation is somewhat different in the second case, when
gluino is lighter than squark, which is the case of the EGRET
point.  Here the expression for the width is more complicated
to be written explicitly~\cite{decay}; however, one can
examine the essential part. When gluino is almost on the mass
shell, a contribution to the amplitude is proportional to
\begin{equation}
\label{wid}
\Gamma(\tilde{g}\rightarrow\tilde{q }q\chi_{0}^{2}))\sim
\frac{3-body\ phase\ space}{(m_{\tilde{q}}^2-(m_{\tilde{g}}
+ m_{q})^2)^2}.
\end{equation}

One can see that the bracket in the denominator which comes
from the squark propagator is minimized for the lightest
squark, since contrary to the previous case gluino is light.
Thus it gives the largest contribution for the third
generation for the same reason as above. The contribution to
the top quark compared to the bottom one is suppressed due to
the phase space which contains only light particles and is
essentially reduced for the heavy top.

This way one can justify that in jet production $b$-jets are
dominant in both cases, which sounds promising for
experimental observation.

\section{Conclusion}

Thus, we conclude that in the EGRET preferred region of
parameter space of the MSSM (small $m_{1/2} \sim 180$~GeV and
large $m_0\sim 1400$~GeV) characterized by a considerable
splitting between the scalar superpartners and gauginos, the
cross-sections for sparticle production at hadronic machines
may reach a few \%~pb. Their decay modes lead to jets (mostly
$b$) and/or leptons in the final states with additional
missing energy carried by escaping neutralinos. These events
have an exceptional signature (see e.g. Table~\ref{n}) that
allows one to distinguish them from the SM background
providing enough integrated luminosity and may be promising
for SUSY searches within the advocated scenario.

\subsection*{Acknowledgements}
Financial support from RFBR grant \# 05-02-17603, grant of
the Ministry of Science
and Technology Policy of the Russian Federation \# 5362.2006.2,
DFG grant \mbox{\# 436 RUS 113/626/0-1} and
the Heisenberg-Landau Programme are kindly
acknowledged. We would like to
thank W. de Boer, V. Zhukov, and V. Bednyakov for valuable
discussions, and A. Pukhov for providing us with
the latest version of CalcHEP and for numerical consultations.

\appendix

\section{Cross-section of a cascade decay}

Calculation of cross sections for complex
cascade processes is a complicated task if done
directly. To simplify it, one can use the
following simple trick: one calculates the
cross-section of the original $2\rightarrow2$
process and then just multiplies it by the
branching ratios for the corresponding unstable
particles at each
junction~\cite{Bilenky}.

\begin{figure}[b]
\centerline{\psfig{file=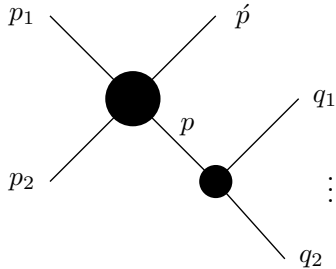,width=0.35\textwidth}}
\vspace*{8pt}
\caption{The cascade process with unstable particle decay. \label{samp}}
\end{figure}

To justify this algorithm, we demonstrate below
how it works for a sample process shown
in Fig.\ref{samp}. We have an interaction of
two particles with  momenta $p_1$ and $p_2$
producing two other particles with  momenta
$p'$ and $p$, respectively.  The particle
with momentum $p$ is unstable and decays into
two particles with momenta $q_1$ and $q_2$.

The cross section of this process is given by
\begin{equation}
d\sigma = \frac{1}{2E_12E_2|v_1-v_2|}
\left(\prod_f \frac{d^3p_f}{(2\pi)^32E_f}\right)
|M|^2 (2\pi)^4\delta^{4}(p_1+p_2-\sum p_f).
\end{equation}
The matrix element for this kind of a process looks like
\begin{equation}
M\sim\bar{u}(q_1)\bar{u}(q_2)V_2D(p)\bar{u}(p')V_1
u(p_1)u(p_2)\delta^4(q_1+q_2-p)d^4p,
\end{equation}
where $u(p)$ corresponds to the external particle and equals  $1,
u_\alpha(p)$ and $\epsilon_\mu$ for a scalar, spinor and vector
particle, respectively, and $D(p)$ is a propagator of an
unstable particle with momentum
$p$ which we take in the form
$$
D(p)=\frac{s(p)}{p^2-m^2+im\Gamma} $$ Here $\Gamma$
is the total width of an unstable
particle and $s(p)$ is a spin dependent part.

To calculate the cross-section, one has to take
the square of the matrix element and
evaluate the phase space integral. One has
\begin{eqnarray}
&&d\sigma \sim \nonumber \\[2mm]
&&\frac{1}{(2\pi)^5}|\bar{u}(q_1)\bar{u}(q_2)V_2D(p)\bar{u}(p')V_1
u(p_1)u(p_2)\bar{u}(p_2)\bar{u}(p_1)V^*_1u(p')D^*(l)V^*_2u(q_1)u(q_2)|\nonumber\\
&&\times \ \delta^4(q_1+q_2-p)\delta^4(q_1+q_2-l)d^4pd^4l  \delta^4(p_1+p_2-p'-q_1-q_2)
\frac{d^3q_1d^3q_2d^3p'}{2E_{q1}2E_{q2}2E_{p'}}\nonumber\\
&&=\frac{1}{(2\pi)^5}\frac{|\bar{u}(q_1)\bar{u}(q_2)V_2s(p)\bar{u}(p')V_1
u(p_1)u(p_2)|^2}{(p^2-m^2)^2+m^2\Gamma^2}\delta^4(p-q_1-q_2)d^4p
\nonumber \\
&&\times \ \delta^4(p_1+p_2-p'-p)\frac{d^3q_1d^3q_2d^3p'}{2E_{q1}2E_{q2}2E_{p'}}
\label{sig}
\end{eqnarray}

When  $\Gamma\ll m$, which means that the particle
is relatively stable and is almost on
mass shell, one can use the approximate formula
\begin{equation}
\frac{1}{(p^2-m^2)^2+m^2\Gamma^2}\approx\frac{\pi}{m\Gamma}\delta(p^{2}-m^{2}).
\end{equation}
This is the first step of approximation. Substituting it into eq.(\ref{sig}) one gets
\begin{eqnarray}
d\sigma&\sim&\frac{1}{(2\pi)^5}|\bar{u}(q_1)\bar{u}(q_2)V_2s(p)\bar{u}(p')V_1
u(p_1)u(p_2)|^2\frac{\pi}{m\Gamma}\delta(p^{2}-m^{2})\nonumber \\
&\times&\delta^4(p-q_1-q_2)\delta^4(p_1+p_2-p'-p)d^4p
\frac{d^3q_1d^3q_2d^3p'}{2E_{q1}2E_{q2}2E_{p'}}\label{sig2}
\end{eqnarray}
Now one can use the relation
$$ \delta(p^2-m^2)d^{4}p=\frac{d^3p}{2p_0},$$
that gives
\begin{eqnarray}
d\sigma&\sim&\frac{1}{(2\pi)^5}|\bar{u}(q_1)\bar{u}(q_2)V_2s(p)\bar{u}(p')V_1
u(p_1)u(p_2)|^2\frac{\pi}{m\Gamma}\nonumber \\
&\times&\delta^4(p_1+p_2-p'-p)\frac{d^3p\:d^3p'}{2E_p2E_{p'}}
\delta^4(p-q_1-q_2)\frac{d^3q_1d^3q_2}{2E_{q1}2E_{q2}}\label{sig3}
\end{eqnarray}
Now we have to transform the matrix element.
It is factorized exactly in the case of a
scalar intermediate unstable particle, while
in the case of a spinor or vector particle,
factorization holds only approximately.  One has
$$|\bar{u}(q_1)\bar{u}(q_2)V_2s(p)\bar{u}(p')V_1 u(p_1)u(p_2)|^2\!\!\approx
|\bar{u}(q_1)\bar{u}(q_2)V_2u(p)|^2 |\bar{u}(p')\bar{u}(p)V_1 u(p_1)u(p_2)|^2 $$

This is the second step of approximation. Thus,
the cross-section is factorized into two parts
\begin{equation}
d\sigma_1\sim\frac{1}{(2\pi)^2}
|\bar{u}(p')\bar{u}(p)V_1u(p_{1})u(p_2)|^{2}
\delta^4(p_{1}+p_{2}-p'-p)\frac{d^{3}p\:
d^{3}p'}{2E_{p}2E_{p'}}
\end{equation}and
\begin{equation}
d\Gamma = \frac{1}{(2\pi)^2}|\bar{u}(q_1)\bar{u}(q_2)V_2u(p)|^2\delta^4(p-q_1-q_2)
\frac{d^3q_1 d^3q_2}{2E_{p}2E_{q1}2E_{q2}},
\end{equation}
so that one can write
\begin{equation}
d\sigma=d\sigma_{1}d\Gamma\frac{E_p}{m\Gamma}.
\end{equation}
Remind now that for a decaying particle
with momentum $p$, its lifetime is given by:
$$\tau=\frac{1}{\ \Gamma_{tot}}=\frac{E_p}{m\Gamma}.$$
This finally gives the desired relation
\begin{equation}
d\sigma=d\sigma_{1}\frac{d\Gamma}{\ \Gamma_{tot}}.
\end{equation}

Thus, we finally come to the conclusion
that for cascade decays the total cross-section
can be obtained by multiplying the original
two-particle one by the branching ratios for
each unstable particle.

\end{document}